# Evaluating Students' Perspectives on ICT Readiness in Somali Higher Education towards Teaching - Learning Acceptance

[1]Yunis Ali Ahmed1,[2]Mohamed M. Mohamed,[3]Abdifatah Farah Ali,[4]Mohamud M. Alasso,[5]Ahmed Dahir Siyad [6]Mohammad Nazir Ahmad

[1,2,34,5]Faculty of Computing, SIMAD UNIVERSITY, Mogadishu, Somalia

[6]Institute of Visual Informatics, Universiti Kebangsaan Malaysia (UKM), 43650 Bangi, Malaysia

E-mail:

[1]yunisali@simad.edu.so,[2]myare81@simad.edu.so,[3]fitaaxfarah@simad.edu.so.[4]alasow@simad.edu.so

[5]siyad@somaliren.org [6]mnazir@ukm.edu.my

## Abstract

Along the rapid development of Information and communication technology (ICT) tools and growth of Internet access offer opportunities that facilitate teaching and learning activities in the context of higher education. However, the study of ICTs readiness and acceptance in Somalia higher education is meagre. This research aims to examine the current state of ICT readiness among university students and explores the factors that affect their readiness acceptance. It proposes an extended model, based on the Technology Acceptance Model (TAM), which explains how University students' beliefs influence their readiness to accept ICT applications in their learning. Survey responses of 304 students from undergraduate and Graduate in Somalia higher education were collected and analyzed using structural equation modelling. The results of the data analysis demonstrated that the TAM explained university students' readiness acceptance of ICT applications reasonably well. More specifically, perceived usefulness, Ease of Use, ICT Self-efficacy, Teaching-Learning autonomy, Student's Optimism and Availability of ICT infrastructure are robust predictors of Students ICT readiness acceptance. Results also showed that internet affordability, network speed and quality, innovativeness, discomfort and insecurity do not have a meaningful effect on perceived usefulness and Ease of Use towards ICT readiness acceptance. Through the empirical results, this study helped us understand why students choose to engage in ICT applications for their learning context.

**Keywords**: ICT readiness acceptance, Higher education,Teaching- Learning, Technology Acceptance Model

## Introduction

Living in the information age, the rapid development of information and communication technology (ICT) has played an important role in various areas, which has brought upon remarkable changes in these areas as well as a lot of positive impact on almost every aspect of human life such as education, health, business, and government (Buabeng-Andoh & Yidana, 2014; Koç, Turan, & Okursoy, 2016; Liu, Toki, & Pange, 2014). ICT has been defined as "Technological devices (hardware and software) that allow to edit, produce, store, exchange and transmit data between different information systems that have common protocols. These applications, which integrate computer media, telecommunications and networks, enable both interpersonal (person to person) and multidirectional (one to many or many to many) communication and collaboration. These tools play a substantive role in the generation, exchange, diffusion, management and access to knowledge"(Romaní, 2009, p1). ICT offers a diverse set of important technological tools and applications which has facilitated information exchange, its storage and dissemination all over the globe, including in developing countries (Copriady, 2014). Notably, the use of advances of ICTs such as laptop computers, data projectors, Smartboards and other digital devices in higher education institutions have shifted the teaching and learning methods from the traditional approaches to a more technologically driven, modern approach (Wu, Pan, & Yuan, 2017). In view of this, use of ICT tools as pedagogical approaches in Somali higher institutions classrooms remains widely erratic, a given potential benefits and promises of greater flexibility teaching and learning outcomes, only a limited number of existing system features are in use(Omer et al., 2015). Further, in East Africa readiness and adoption of ICTs technologies in higher education remains challenges (Tulinayo, Ssentume, & Najjuma, 2018).

Globally, huge investments in ICT infrastructure and the training of educators have been initiated by many countries to promote the appropriate use of ICT applications in educational institutions. Some examples are, the USA in 2015, educational institutions spent about $6.6 billion on ICT resources such initiative networks, notebook and learning apps to improve student learning entire (McCandless, 2015). In 2009, the United Kingdom spent £2.5 billion to promote use of ICT in higher institutions (Nutt, 2010). In 2013 and 2014, the annual budget allocated for improving ICT infrastructure in Taiwan's higher education sector averaged at US$5.5 million (Wu et al., 2017). These constitute interventions are aimed to enhance ICT integration entire education system. As a context of this research, Somalia since 2000 there has been increasing a big push to introduce digital technology include computer hardware, software and internet services, into higher education institutions supported by World Bank (Hare, 2007). Nevertheless, all challenges aspect to implementations of ICT applications must be "ready" such technology, learner and instructors to build up a coherent communication(Mosa, Mahrin, & Ibrrahim, 2016) as nature of ICT it enables new ''possibilities'' for learners; it does not provide a ''ready to use'' resource(Sørebø, Halvari, Gulli, & Kristiansen, 2009).

The use of ICT in developing countries including Africa has been a significant increase in recently years, nevertheless, a numerous problems encountered by higher education intuitions such low ICT literacy levels, insufficient resources, poor ICT infrastructures, poor internet connection, limited fund and support from government(Omer et al., 2015). Additionally, Somalia has had a traumatic past experienced by violence, war and instability since 1991 which entirely effected all public social services including education systems, Notwithstanding, in last two decades

Significant progress in post-conflict rebuilding that has resulted in the rapid proliferation of training colleges and universities  has been occurred(Eno, Eno, & Mweseli, 2015). Somalia has two categories of higher learning institutions: Public institutions managed by the federal government and owned Private intuitions. Despite the protracted conflict and violence in Somalia the computer-based technologies and internet services and  use of  mobile networks with availability of social networking tools is already widespread through the country (Ahmed, Ahmad, & Zakaria, 2016; Cooley & Jones, 2013). In view of this, the study examines student perceptions of readiness and acceptance of ICT applications to use in teaching and learning.

Although Somalia has one of  the lowest penetration rates of ICT applications when compared with other African countries(Souter & Kelly, 2013), the most of these higher learning institutions have been performing their everyday operations through emerging ICT technologies without a formal setting. For instance, beginning from the introduction of the ICT applications and internet to the Somali higher institution, staff and university students have been utilized  for academic purposes, where students are expected to enhance their learning capabilities such managing courses material    and programs to support the traditional teaching and learning delivery system(Omer et al., 2015). Also, ownership of Smartphones, computer laptops have been increasing among the university students. However, in Somalia, the use of ICT tools for educational activities is something which is still in the early stages of development. Due to its newness, no literature is available concerning the readiness of new learning technologies in higher educations of leaning in Somalia. Furthermore, no research studies have been conducted so far with the aim of evaluating students 'readiness and acceptance of ICT tools for teaching and learning purposes. Findings of previous higher education studies stated that a lack of ICT readiness and acceptance hampers the realization of its benefits (Ayele & Birhanie, 2018; Ghosh Roy & Upadhyay, 2017). Similar study "the challenge for academia is to stay abreast or ahead of student technology acceptance"(Apostolou, Dorminey, Hassell, & Watson, 2013). In this sense, there is a need to evaluate students' varying perceptions and levels of readiness to acceptance of ICT application tools in teaching and learning perspective will have paramount importance.

In response, this study aims to fill this gap in the literature and address these problems by empirically examining the main determinants of ICT readiness acceptance and proposing a research model that incorporates the determinants and aspects for ICT enablers in terms of organizational readiness (i.e. ICT infrastructure, network speed and quality and internet access and affordability), personal readiness (i.e. optimism, innovativeness, discomfort, and insecurity) and individual control beliefs (i.e.  Learning autonomy and ICT Self-efficacy). The study investigates how these identified key drivers influence among the students' ICT readiness acceptance process within higher educational institutions. To guide this research effort, we conceptualized extended theoretical framework of technology acceptance model (TAM) proposed by (Davis, 1989). TAM is a simple and considered as a robust framework and has been utilized extensively and successfully to explain and predict the influence of technology on user behavior. The original TAM model applies  factors  perceived usefulness and perceived ease of use  which explains why users decide to accept or reject use of particular technology(Davis, 1989). Nevertheless, numerous authors criticized that TAM need "for additional variables to enhance its predictive power"(Granić & Marangunić, 2019; Venkatesh & Davis, 2000). In this study it is attempted to introduce an integrated a new ICT readiness acceptance model by extending the TAM model by adding constructs derived from literature to provide useful insights into the students' readiness to accept ICT tools in the field of learning and teaching in Somalia.

A total of 304 questionnaires were used to collect data from students of the 5 largest public and private higher educational institutions in Somalia. To evaluate the research model, this study applies structural equation modeling (SEM). The findings of our paper make contributions which are twofold: First, in this study seeks to add new knowledge to the ICT readiness acceptance literature in higher education by exploring from a multi-faceted viewpoint including organizational readiness, personal readiness and control beliefs towards ICT readiness. Second, since, the readiness and acceptance of new technology in teaching-learning courses among the university students is still at its infancy, in this study builds theoretical framework that support acceptance of ICT learning in east Africa countries particular in Somali higher institutions. This paper also helps researchers and practitioners to develop approaches for improving students 'self-readiness awareness.

The rest of this paper is organized as follows. In Section 2, we present our theoretical framework and hypotheses Development. Then in Section 3 discusses the research methodological procedures. In section 4 we explore the findings of our study. In the final section, discussions, conclusions, limitations and future research of ICT readiness acceptance are presented.

## Theoretical framework

## Acceptance of ICT readiness in higher educational settings

As the positive effects of ICTs in higher Institutions provides students and lectures advanced skills which has led to their learning to be more dynamic and accessible(Al-Adwan, Al-Madadha, & Zvirzdinaite, 2018). ICT applications are an "all-encompassing term that includes the full gamut of electronic tools by means of which we gather, record and store information, and by means of which we exchange and distribute information to others" (Anderson, 2010; p.4). During recent decades, the widespread use of modern ICT applications present a variety of advantages for learners, course, and teacher, for example, ICT provide university students learning platform designs that provide a pathway to improve student learning processes(Vega-Hernández, Patino-Alonso, & Galindo-Villardón, 2018). Accordingly, introduction of ICT applications and integration in education context has been seen as important and the increasing its demand in low income counties of Africa to embrace these technologies for teaching and learning(Chitiyo & Harmon, 2009). However, today, the first vital step of use ICT applications in higher education is user' readiness and acceptance or reject new ICTs (Gombachika & Khangamwa, 2013), Such readiness can be described as how ready the higher institutions is on several aspects of technology to implement into teaching and learning(Schreurs, Ehlers, & Sammour, 2008), thus, exploring the learners' perceptions and understanding the level of readiness of their institutions and the intended users is necessary (Dray, Lowenthal, Miszkiewicz, Ruiz-Primo, & Marczynski, 2011).

In the literature, researchers found that the testing the silent factors and causal relationships that influence individuals and institutions to participate in the utilization of ICT for teaching and learning (Chipembele & Bwalya, 2016). There are many factors involved in such an initiative; for example, a study conducted by Dumpit and Fernandez (2017) developed an extension of TAM model to explain the factors that determining adoption of emerged ICT initiative among the students in university. The study found that factors related in ICT infrastructure such internet reliability and speed and quality were only significant public higher institutions. Another study by Basri, Dominic, and Khan (2011) suggested a research framework which is composed of organizational-readiness factors, and another relevant framework that effected readiness of e-technologies (i.e. e-maintenance, e-procurement and e-learning) adoption. Similarly, Aboelmaged, (2014) proposed environmental, technological and organizational factors that examined ICT readiness, such as e-maintenance technology readiness adoption in manufacturing firms. Within the Information Systems (IS) literature a considerable amount of past studies have used technology acceptance and adoption theories to explain user behavior towards ICT readiness and acceptance in higher institutions including Diffusion of Innovation Theory (Rogers, 1995) ,Theory of Reasoned Action(Fishbein & Ajzen, 1977) and theory of planned behaviour(Ajzen, 1985). However, among these models, the TAM Model is particular interest in this study.

**The TAM Model**

The antecedent of TAM model was based on the Theory of Reasoned Action (TRA)(Fishbein & Ajzen, 1977). The TRA theory take into account person's behavior through behavior intention shaped by attitude and subject norms which model assumes that most behaviors related in the field of social psychology. Davis developed a new simplified model for assessing information System acceptance process(Davis, 1985). The TAM explains why a wide range of end-users may accept or reject new system technologies. For example, previous studies have suggested in the literature that the TAM model has good predictive validity for the usage of ICT applications including e-mail, e-learning, social media usage, web technology and instant messaging (Ahmed et al., 2016; Al-Hawari & Mouakket, 2010; Ngai, Poon, & Chan, 2007)."A key purpose of TAM is to provide a basis for tracing the impact of external factors on internal beliefs, attitudes, and intentions"(Davis, 1985). TAM suggests that two beliefs— perceived usefulness (PU) and the perceived ease of use (PEO) which are primary determinants of an individual's intention to acceptance and use of the technology (Davis, 1989).

The TAM has been used to evaluate use of technology term systems and interaction in higher education, the user can decide whether the ICT tools are useful or easy to use. Specifically, in the TAM suggests that two beliefs— perceived usefulness (PU) and the perceived ease of use (PEOU) have been shown to affect the use of ICT tools in higher education (Elwood, Changchit, & Cutshall, 2006; Koç et al., 2016). The PU has been defined as the prospective users' beliefs towards use of particular systems technology will improve her/his productivity or work performance(Davis, 1989). Previous research conducted in educational field provided evidence that TAM as tool used to understand student's perceptions about how PU and PE effect their decision on readiness and acceptance of the system (Park, Nam, & Cha, 2012). Although, TAM model proposed by Davis present flexible, simplest which able to explain

**Research model and hypotheses development**

The Technology Acceptance Model (TAM) explains why a wide range of end-users may accept or reject new technologies. For example, previous studies have suggested in the literature that the TAM model has good predictive validity for the usage of ICT applications including e-mail, e-learning, social media usage, web technology and instant messaging (Ahmed et al., 2016; Al-Hawari & Mouakket, 2010; Ngai et al., 2007). Based on the context of use of technology and interaction in higher education, the user can decide whether the ICT tools are useful or easy to use. Specifically, in the TAM, two beliefs— perceived usefulness (PU) and the perceived ease of use (PEOU) have been shown to affect the use of ICT tools in higher education (Koç et al., 2016). PU and PEOU are hypothesized for their importance and are associated with technology in various ways. Both of these beliefs are considered as fundamental determinants of user perception and acceptance of a given ICT component (Sanchez-Franco, 2010).Thus, in this study the first two hypotheses (H1–H2) originate from existing literature on TAM (Bhattacherjee, 2001; Venkatesh, Morris, Davis, & Davis, 2003). This leads to the following hypotheses:

**H1**. Perceived usefulness positively influences ICT readiness acceptance

**H2.** Perceived ease of use positively influences ICT readiness acceptance

**Organizational readiness**

Technology Infrastructure refers to the availability of a requisite technology platform that facilitates and provides solutions throughout an organization (Bhattacherjee & Hikmet, 2008). Availability of a well-designed infrastructure that is composed of all the necessary ICT applications (i.e. computer hardware, software, and networking technologies) indicates potential technology readiness efforts (Aboelmaged, 2014), which forms the basis for the promotion of an e-learning platform infrastructure, and hence leads to improvements in teaching and learning activities (Lwoga, 2012). In contrast without an existing infrastructure in terms of ICT functions, an organization or individual may limit their technology choices and usage (Bhattacherjee & Hikmet, 2008).

Reflecting existing literature, the availability of infrastructure in organizations (i.e. higher education sector) can increased their readiness to use such systems. Several studies have tried to explain organizational readiness, including availability infrastructure and its usefulness, and have related these to the successful implementation of such enabling e-governance services (Ghosh Roy & Upadhyay, 2017), university electronic records and learning platforms (Eze Asogwa, 2013) and health information systems (Bhattacherjee & Hikmet, 2008). Accordingly, this study hypothesizes that the availability of an ICT infrastructure will increase organizational perceived usefulness of new systems.

**H3.**  Availability of infrastructure positively affects user's perceived usefulness

Access to internet technologies is becoming a necessity not only for gaining information and communication, but also for education and health. The Internet has become the core of virtually every aspect of our daily lives, and an integral element of education for most people, whereby learning and teaching through online communication has made it easy for people (Ramani, 2015; Sagan & Leighton, 2010). Offering affordable internet services with minimal costs has been documented, particularly linking income of learners towards affordability of technology (van Mourik, Cameron, Ewen, & Laing, 2010; Weiss, Gulati, Yates, & Yates, 2015). The International Telecommunication Union (ITU) has been defined Internet affordability as "the price of broadband Internet service and cost of devices relative to income" (ITU, 2016).

Studying of Internet affordability is of prime importance because it means measuring the extent to which internet bandwidth costs as a percentage to the total expenditure of the institution or campus, and determining whether it is affordable; understanding the users perceptions towards affordability of these technologies, particular in higher education setting is an emerging area of research (Echezona & Ugwuanyi, 2010; Penard, Poussing, Mukoko, & Piaptie, 2015; Yates, Gulati, & Weiss, 2010). Thus, this study hypothesizes that Internet affordability has an influence over user's perceived usefulness of internet access from the teaching and learning prospective.

**H4.**  Internet affordability positively affects user's perceived usefulness

Speed and quality of the network and internet connection within the organization were the most important factors to be looked into. The amount of throughput which the users can get between two devices has seen a general improvement (Simsim, 2011). A problem worth exploring is, how to share internet resources within an organization with limited available bandwidth, in particular in higher educational institutions. The introduction of computers with high speed networks based on teaching-learning approaches has led to important innovations in educational content (Li, 2010).

Within the context of higher education, speed and quality of internet networks are the twin weapons which may support teaching and learning activities through social technologies (Chai & Lim, 2011; Matthews & Schrum, 2003). Access to a high speed network with sufficient internet bandwidth also helps in interactions among educational administrators by allowing them to expand the teaching-learning contents in their institutional workplaces (Matthews & Schrum, 2003). Hnce, this study hypothesizes the following:

**H5.** Network speed and quality positively affect the perceived usefulness towards ICT readiness acceptance

**Personal readiness**

Several attempts have been made to understand enablers or exhibitors that influence a person's tendency to embrace and use new technologies (Nugroho & Fajar, 2017; Son & Han, 2011; Svendsen, Johnsen, Almås-Sørensen, & Vittersø, 2013). Gombachika and Khangamwa, (2012) postulate that personality traits play a role in determining the user's predisposition in the adoption of ICT readiness in education. Personal readiness is defined as the willingness of an individual to try out any new information technology (Nov & Ye, 2008). Through the development of the Technology Readiness Index (TRI) proposed by (Parasuraman, 2000), four personal trait constructs are presented, which are optimism, innovativeness, discomfort and insecurity. These determine the readiness of individuals to use new technologies (ICT tools) for achieving personal- and work-related goals.

The personal optimism construct relates to a positive view of technology to be used and a belief about whether that technology will increase control, flexibility and efficiency of the prospected users (Parasuraman, 2000). Period researches argue that a high personal optimism (motivation) to adopt ICT readiness in institutions and a belief in ICT applications improve their control, flexibility and user activities efficiently (Gombachika & Khangamwa, 2012). Literature in the field of ICT readiness acceptance has also highlighted that optimism is significantly related to perceived usefulness and the ease to use ICT tools. Hence, this study hypothesizes the following:

**H6a.** Personal optimism positively affects the PU towards ICT readiness acceptance

**H6b**. Personal optimism positively affects the PEOU towards ICT readiness acceptance

In general innovation diffusion research, Personal Innovativeness represents as the tendency to be the first to use a new technology and a tendency to be believed leader(Parasuraman, 2000), and it has long been recognized as an important determinant of individuals with high technological innovativeness which in turn, influences PEOU and PU (Erdoğmuş & Esen, 2011; Yi, Tung, & Wu, 2003). In a higher education context, lecturers and learners that have perceived high innovativeness are active information seekers about new ideas and they develop more positive technology acceptance (Svendsen et al., 2013). Hence, this study proposes the following hypotheses:

**H7a.** Personal Innovativeness positively affects the PU towards ICT readiness acceptance

**H7b**. Personal Innovativeness positively affects the PEOU towards ICT readiness acceptance

Personal Discomfort is defined as "a perception of being unable to control the technology and a feeling of being overwhelmed by it" (Parasuraman, 2000). For instance, individuals who are uncomfortable with the use of new technologies believe that they are controlled by it (Parasuraman, 2000; Walzuch, Lemmink, & Streukens, 2007). This construct is posited to influence user control and a sense of being overwhelmed. The low self-esteem experienced in using the latest ICT applications has been indicated to have an impact on user's perceived ease of use and usefulness. Thus, we propose the following hypotheses:

**H8a.** Personal Discomfort positively affects the PU towards ICT readiness acceptance

**H8b**. Personal Discomfort positively affects the PEOU towards ICT readiness acceptance

In this study, Insecurity is described as a scenario when people distrust technology for security and privacy reasons (Parasuraman, 2000). The characteristics of security such as confidentiality, integrity and availability can contribute to the degree of users' positive perception and beliefs from the technology (Lu, Yao, & Yu, 2005). Based on period studies, insecurity is regarded as an "inhibitor", which can improve the level of individual's readiness in using the technology, which in turn will lead to a higher rate of use of ICT tools in higher educational institutions. (Al-Hawari & Mouakket, 2010) confirmed that the feelings and perception of insecurity influence PU and PEOU (Kuo, Liu, & Ma, 2013). Thus, we suggest the following hypotheses:

**H9a.** Personal Insecurity positively affects the PU towards ICT readiness acceptance

**H9b**. Personal Insecurity positively affects the PEOU towards ICT readiness acceptance

**Control beliefs towards ICT**

Studying of the control beliefs about learning and instruction in a higher education context, recent studies reported that the teachers' and leaners' beliefs can be barriers to ICT integration (Klassen, Tze, Betts, & Gordon, 2011). For instance, learners' opinions about the ease or difficulty of use of ICT in higher educational institutions can be termed as individual control beliefs. According to Ajzen (1991), a control belief is defined as the belief that predicts about the presence of factors and their influence effectively in a given situation (outcome evaluations). This paper examines the control beliefs that self-efficacy and learning autonomy contribute to the ease of use of ICT applications in higher educational institutions. Learning autonomy is referred to as the learners' control over learning the process with advanced ICT applications, which provides a more dynamic approach, reflecting on how they achieve their individual learning goals (Fernandez, 2000). Previous studies found that learning autonomy has become an increasingly important part of ICT readiness in higher education by encouraging the learner to become a main contributor to system usage and acceptance (Cheon, Lee, Crooks, & Song, 2012). Although ICT tools support interactivity in a variety of teaching and learning contexts, and could provide more flexibility, learning autonomy requires an understanding of learners' readiness behaviors (Nakata, 2011). Thus, autonomy is an important factor of control beliefs for ICT readiness. Accordingly, this study develops the following hypothesis:

**H10.** Perceived learning autonomy positively affects the perceived ease of use towards ICT readiness acceptance

Self-efficacy refers to ''people's beliefs about their capabilities to exercise control over their own level of functioning and over events that affect their lives'' (Bandura, 1991, p. 257). This study of ICT self-efficacy in the context of higher education is concerned with the convictions an individual has for their ability to use a set of tools enabling and supporting teaching and learning activities, and how learners can overcome the impediments that hamper the use of ICT technologies. Self-efficacy is the key determinant of coping behavior of self-motivation, which can help understand how ICT in education influences someone's choice to be engaged (digital native students, teachers' changing role) in teaching and learning effectiveness (Sang, Valcke, Van Braak, & Tondeur, 2010). Based on this the following research hypothesis is proposed:

**H11.** ICT Self-efficacy positively affects the perceived ease of use towards ICT readiness acceptance

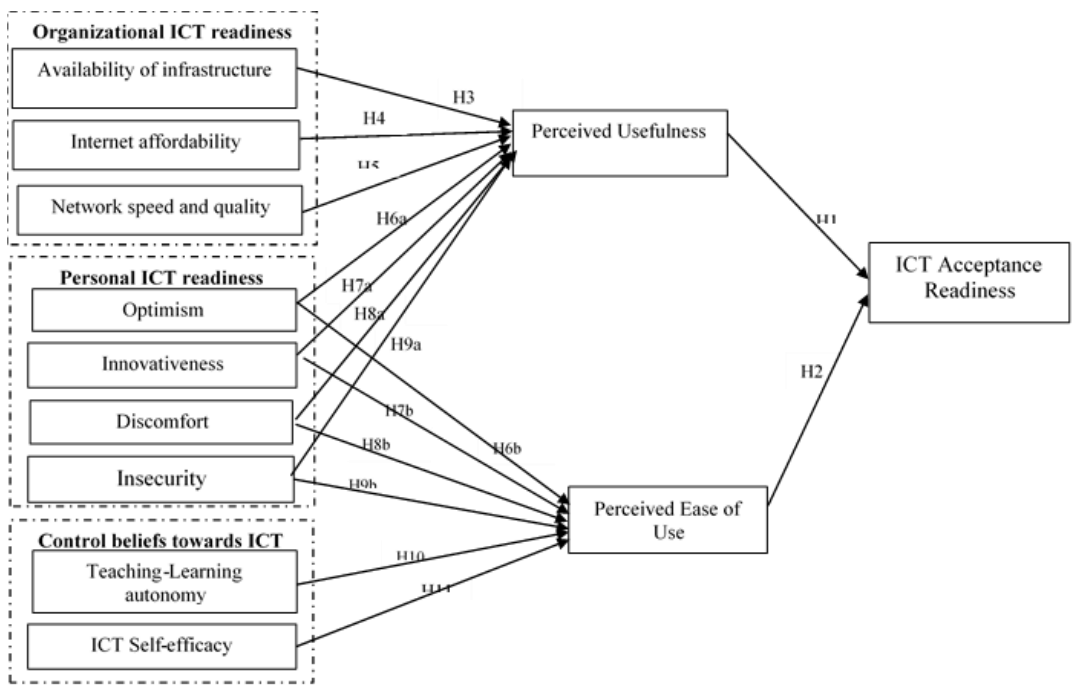

Figure 1: Conceptual model of Research

**Research Method**

**Participants and sample**

A survey research was conducted using a pre-tested questionnaire at Somali private and public Universities. This study employed a non-probability sampling method (i.e., purposive sampling) ) to carry out  data collection (Creswell, 2002) through the targeted respondents in this study, which were heterogeneous senior students from universities belonging to different faculties who had voluntarily agreed to participate. A total of 304 questionnaires were distributed throughout 5 of the largest universities in Somalia. 304 questionnaires were completed and returned, representing a response rate of 91%, because 28 participants were removed due to missing responses. The questionnaire comprised of two parts: Part one focused upon the respondents' demographic data, while part two concentrated primarily upon ICT readiness acceptance and identifying the factors that could contribute to ICT usage in Somali higher educational institutions.

**Measurement development**

To test the hypotheses in our study, we collected data from private and public universities in Somalia, utilizing a questionnaire survey approach to determine the ICT readiness in Somali higher education institutions. The study selected 5 largest universities with student populations of more than 3000 and which were established prior to 2009, and were using ICT applications as their teaching-learning platform. The questionnaire was pre-tested in the field before the main data collection. The survey instrument contained 50 items from 12 different constricts adopted from previous literature. Due to the lack of context-specific period literature, in this study several instrument items have been edited and reworded. The study measured participants' perception towards ICT readiness acceptance using a 5-point Likert scales due to their ability to balance detailed about the respondents choice for selecting an appropriate survey questions(Weijters, Cabooter, & Schillewaert, 2010) The responses ranged from strongly disagree (1) to strongly agree(5).

**Data Analysis**

Construct validity was used to validate the instrument. To test the research model shown in Fig. 1, in this study, the Partial Least Squares (PLS) approach using the SmartPLS 3.0 software package was employed to analyses data and to evaluate the research model. This software is capable of handling non-normal data and providing the end-all solution to models (F. Hair Jr, Sarstedt, Hopkins, & G. Kuppelwieser, 2014; J. Hair, Hollingsworth, Randolph, & Chong, 2017). Another determinant of this was the small sample size that can influence activity aspects of SEM; for instance, applying PLS SEM to the model estimation is typically dealing with much smaller sample sizes, without considering whether the model is complex or not (Chin, 1998; Joe F Hair, Sarstedt, Ringle, & Mena, 2012). SmartPLS software package V3.2.7 was employed to analyze data and to evaluate the research model.

From the validation and testing of the research model of this study, both, the measurement model and the structural model were estimated. To evaluate the measurement model the study tested construct validity (Convergent and Discriminant validity). The convergent validity was used to assess three standard conditions of validity; (1) internal consistency (Composite Reliability CR), (2) indicator reliability (indicator factor loadings) and (3) convergent validity (AVE) respectively(Bagozzi & Yi, 2012). As per recommendations, the indicator loadings should exceed 0.5 (Joseph F Hair, Anderson, Tatham, & Black, 1998), CR should exceed the minimum of 0.7 (Joseph F Hair, Ringle, & Sarstedt, 2013), and the AVE of every construct should be more than 50 percent of variance (Fornell & Larcker, 1981). Due to their poor value of scales, some items were omitted during this validation process, while validating to the Composite Reliability's values or when the factor loadings were weak, with loadings less than 0.5. Reliability analysis is achieved when the CR exceeds the minimum of 0.7 (Marôco, 2010).

**Results**

**Respondent demographics**

Table 1 presents the profile of the respondents. Males (79.6%) were slightly more than the females (20.4%) which somewhat reflects the gender ratio of the students in Somali higher education institutions. Most of the respondents were aged between 21-25 (47.7%). This is because an increasingly higher number of young students from secondary schools have joined higher education in the last few years. About 82.2% of students are undergraduate level. In this study, it was also found that 36.8% of the students used ICT applications for 6–10 h daily, 21.4% of the students used ICT applications for 1–5 h daily, whereas 20.4% of the students used ICT applications for 11–15 h or more, daily.

Table 1: Profile of Respondents

| Profile | | Frequency | Percentage (%) |
|---|---|---|---|
| Gender | Male | 242 | 79.6 |
| | Female | 62 | 20.4 |
| Age | Below 20 | 39 | 12.8 |
| | 21-25 | 145 | 47.7 |
| | 26-30 | 84 | 27.6 |
| | 31-40 | 27 | 8.9 |
| | Above 40 | 9 | 3.0 |
| Group | Graduate | 54 | 17.8 |
| | Undergraduate | 250 | 82.2 |
| Time spent on the ICT daily | < 1 h | 38 | 12.5 |
| | 1–5 h | 65 | 21.4 |
| | 6–10 h | 112 | 36.8 |
| | 11–15 h | 62 | 20.4 |
| | More than 15 h | 27 | 8.9 |

**Assessment of Measurement model**

The measurement model tested two types of validity: convergent and discriminant validity analyses. Based on the data analysis result convergent validity was achieved as seen in the factor loadings of the items, which exhibited loadings above 0.6, whereas the recommended value of convergent validity should be greater than 0.50. All composite reliability values also exceeded the 0.70 suggested guideline (Hair Jr, Sarstedt, Ringle, & Gudergan, 2017). Table 2 display the summary of the outcome of the analyses.

Table 2. Results for the measurement model

| Construct | Item | Factor Loading | CR | AVE |
|---|---|---|---|---|
| Availability of infrastructure | AVIN01 | 0.832 | | |
| | AVIN02 | 0.755 | 0.822 | 0.607 |
| | AVIN03 | 0.747 | | |
| Internet Affordability | INAF01 | 0.968 | 0.809 | 0.686 |
| | INAF03 | 0.659 | | |
| Network speed and quality | NQ01 | 0.789 | | |
| | NQ02 | 0.707 | 0.797 | 0.567 |
| | NQ03 | 0.761 | | |
| Optimism | OP02 | 0.749 | | |
| | OP03 | 0.661 | 0.784 | 0.549 |
| | OP04 | 0.805 | | |
| Innovativeness | INV01 | 0.853 | | |
| | I INV03 | 0.634 | 0.763 | 0.523 |
| | I INV05 | 0.662 | | |
| Discomfort | DIS03 | 0.828 | 0.829 | 0.709 |
| | DIS05 | 0.855 | | |
| Insecurity | INS03 | 0.794 | 0.811 | 0.683 |
| | INS04 | 0.857 | | |
| ICT Self-efficacy | ICTSE01 | 0.744 | | |
| | ICTSE03 | 0.782 | | |
| | ICTSE06 | 0.614 | 0.758 | 0.514 |
| Teaching and Learning Autonomy | TA01 | 0.719 | | |
| | TA02 | 0.645 | 0.753 | 0.506 |
| | TA03 | 0.765 | | |

| Perceived Usefulness | PU01 | 0.749 | | |
| | PU03 | 0.737 | 0.791 | 0.558 |
| | PU04 | 0.754 | | |
| Perceived Ease of Use | PEU01 | 0.701 | | |
| | PEU04 | 0.742 | 0.777 | 0.537 |
| | PEU05 | 0.755 | | |
| ICT Readiness Acceptance | ICTRA01 | 0.776 | | |
| | ICTRA02 | 0.746 | | |
| | ICTRA03 | 0.687 | 0.781 | 0.544 |

Discriminant validity evaluation is the extent to which a construct is truly distinct from other constructs by empirical standards (Hair Jr et al., 2017). The author highlighted three approaches to assess discriminant validity: (1) Fornell and Larcker criterion, (2) Cross-Loadings, and (3) Heterotrait-Monotrait Ratio (HTMT), which acts as a new criterion for evaluating discriminant validity. The associated information is provided in the Table below, according to Heterotrait-Monotrait Ratio (HTMT) as recommended by (Hair Jr et al., 2017). HTMT could detect the possible in discriminants among the correlations (values) of indicators across constructs. Values close to 1 for HTMT can conclude that there is a problem with the discriminant validity. Some studies suggested a threshold of 0.85 (Kline, 2011) while other authors proposed a value of 0.90 (Gold, Malhotra, & Segars, 2001).

Table 3: Discriminant Validity using Heterotrait-Monotrait Ratio (HTMT)

| | | 1 | 2 | 3 | 4 | 5 | 6 | 7 | 8 | 9 | 10 | 11 |
|---|---|---|---|---|---|---|---|---|---|---|---|---|
| 1 | Avia. Infrastructure | | | | | | | | | | | |
| 2 | Discomfort | 0.094 | | | | | | | | | | |
| 3 | ICT Readiness | 0.592 | 0.295 | | | | | | | | | |
| 4 | ICT Self | 0.323 | 0.225 | 0.631 | | | | | | | | |
| 5 | Innovativeness | 0.230 | 0.175 | 0.430 | 0.514 | | | | | | | |
| 6 | In-Security | 0.313 | 0.369 | 0.392 | 0.410 | 0.305 | | | | | | |
| 7 | Int. Affordability | 0.226 | 0.287 | 0.098 | 0.264 | 0.292 | 0.104 | | | | | |
| 8 | Network. Speed and Quality | 0.361 | 0.440 | 0.361 | 0.299 | 0.274 | 0.556 | 0.482 | | | | |
| 9 | Optimism | 0.576 | 0.146 | 0.529 | 0.481 | 0.602 | 0.406 | 0.198 | 0.0.35 | | | |
| 10 | Perceived Ease of Use | 0.527 | 0.140 | 0.842 | 0.656 | 0.455 | 0.487 | 0.191 | 0.356 | 0.465 | | |
| 11 | Perceived Usefulness | 0.591 | 0.308 | 0.855 | 0.729 | 0.553 | 0.441 | 0.124 | 0.291 | 0.387 | 0.483 | |
| 12 | Teaching and Learning | 0.492 | 0.274 | 0.707 | 0. 84 | 0.478 | 0.487 | 0.319 | 0.268 | 0.581 | 0.505 | 0.612 |

**Assessment of Structural Model**

The structural model is estimated using partial least squares (PLS) V3. The structural model is the second step of SEM, which covers the model's predictive capabilities through R2 values, the goodness of fit model, and also helps in knowing the relationships among hypothetical constructs. If the outer model (measurement model) is reliable, a valid assessment permits an evaluation of the inner path model's (structural model) estimates. The bootstrapping procedure was used to examine the path coefficients, T-statistics values and the significance values by employing 5000 subsamples (Hair Jr et al., 2017).When the absolute value of t is greater than 1.96 then the significance level should be 0.05. Table 4 is presented the results of research hypothesis tested

Table 4 Summary of Hypothesis Testing.

| Hypothesis | Path | Path coefficient(β) | t-value | P value | Results |
|---|---|---|---|---|---|
| **H1** | PU□ICTRA | 0.314 | 5.214 | 0.000*** | Supported |
| **H2** | PEOU□CTRA | 0.404 | 6.999 | 0.000*** | Supported |
| **H3** | AVIN□PU | 0.254 | 5.017 | 0.000*** | Supported |

| H4 | INAF☐PU | -0.031 | 0.609 | 0.543 | **Not Supported** |
|---|---|---|---|---|---|
| H5 | NSQ☐ PU | -0.019 | 0.327 | 0.744 | **Not Supported** |
| **H6a** | OP☐PU | 0.234 | 4.390 | 0.000*** | Supported |
| **H6b** | OP☐PEOU | 0.158 | 2.516 | 0.012** | Supported |
| **H7a** | INV☐ PU | 0.200 | 3.629 | 0.000*** | Supported |
| **H7b** | INV☐PEOU | 0.082 | 1.258 | 0.209 | **Not Supported** |
| **H8a** | DIS☐PU | 0.123 | 2.822 | 0.005** | Supported |
| **H8b** | DIS☐PEOU | -0.023 | 0.446 | 0.656 | **Not Supported** |
| **H9a** | INS☐PU | 0.094 | 1.670 | 0.096 | **Not Supported** |
| **H9b** | INS☐PEOU | 0.127 | 2.304 | 0.022* | Supported |
| **H10** | ICTSE☐PEOU | 0.136 | 2.163 | 0.031* | Supported |
| **H11** | TLA☐PEOU | 0.261 | 3.956 | 0.000*** | Supported |

P < 0.05*, p < 0.01**, p < 0.001 ***

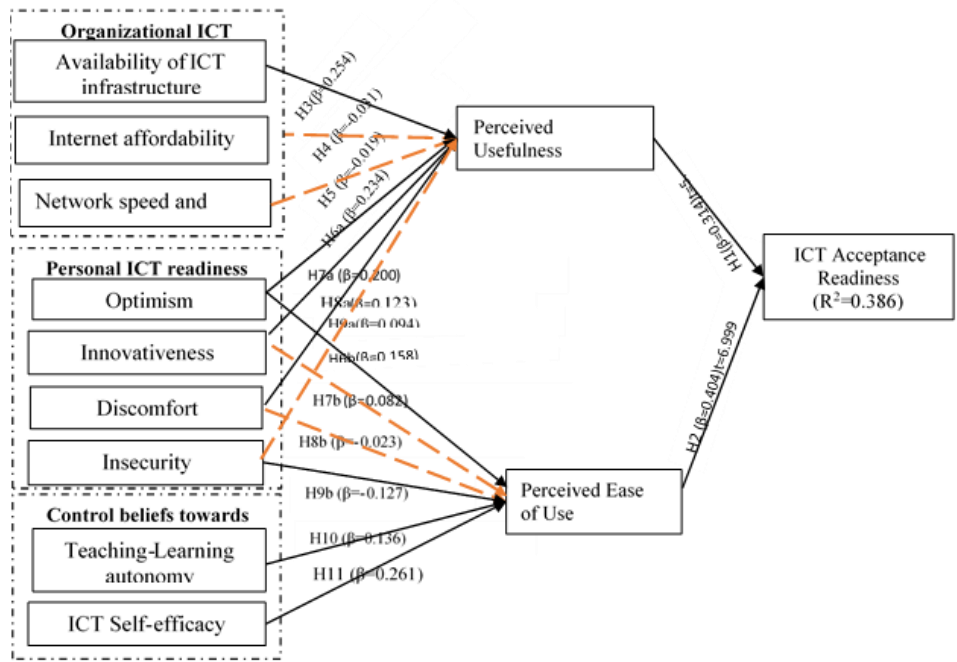

Figure. 2 shows the graphical description of the results of path coefficients

The coefficient of determination ($R^2$) value in the amount of variance of dependent constructs and the antecedents proposed in the research model, which have been evaluated. The table below presents the $R^2$ value of ICT readiness acceptance in higher education, which is noted to be more than 38% ($R^2 = 0.382$). The $R^2$ value ranges from 0 to 1. As a rule of thumb, the values of $R^2$ for independent variables should be equal to or higher than 0.10 (Falk & Miller, 1992).

Table 5: R Square of Endogenous Construct

| Construct | | R2 | B | t Value | P |
|---|---|---|---|---|---|
| ICT acceptance Readiness | PEOU | 0.386 | 0.404 | 6.999 | 0.000 |
| | PU | | 0.314 | 5.21 | 0.000 |

Having tested and evaluated ICT Readiness Acceptance Model via the collected data form the respondents' ICT readiness in Somali higher educations, the two-step sub SEM analysis approaches (e.g. measurement and the structural models) were used. In our theoretical research model of this study, the thirteen latent variables were Cooperation, Availability of Infrastructure, Internet Affordability, Network speed and quality, Optimism, Innovativeness, Discomfort, ICT Self-efficacy, Teaching and Learning Autonomy, Perceived Usefulness, Perceived Ease of Use and ICT Readiness Acceptance. Statistically, as indicated in Table 5, the calculated path coefficients and the hypotheses tested show that six constructs were significant and five hypotheses were insignificant. Out of the six constructs, it was inferred that Perceived Usefulness ($\beta$=.314, p < 0.001,) and Perceived Ease of Use ($\beta$=0.404, p<0.001) have an appreciably positive effect on ICT Readiness Acceptance (H1 and H2 supported).

While checking the coefficients and t statistics between Perceived Usefulness, Perceived Ease of Use and its observed variables, the results showed that Availability of infrastructure ($\beta$=0.254, p<0.001), Optimism ($\beta$=0.234, p<0.001 and $\beta$=0.158, p< 0.01) are positively related, thereby supporting H3, H6a and H6b. As hypothesized in H10 and H11, ICT Self-efficacy ($\beta$=0.136, p < 0.05) and Teaching and Learning Autonomy ($\beta$=00.261, p<0.001) positively influence Perceived Ease of Use. Contrary to expectations, Internet Affordability and Network Speed and Quality was found to have no significant impact on Perceived Usefulness towards ICT Readiness Acceptance, thus rejecting H4 and H5. From the analysis, it can be seen that the hypotheses of Innovativeness (H7a, $\beta$=0.200, p<0.001) and Discomfort (H8a $\beta$=0.123, p < 0.01) have significant influence on Perceived Ease of Use, while H7b and H8b are not significant on Perceived Usefulness towards ICT Readiness Acceptance. There was found to be no significance between Insecurity (H9a) and Perceived Usefulness; on the other hand, Insecurity ($\beta$=0.127, p< 0.05) was found to have a significant influence on Perceived Ease of Use towards ICT Readiness Acceptance. Fig. 2 depicts the path coefficients of hypotheses tested with their significance and the $R^2$-values for each dependent variable included in the ICT readiness acceptance.

## Discussions

The current study has proposed an extended TAM based Model with constructs suitable for developing countries of Africa, such Somalia segment and verified the model in the context of ICTs readiness and acceptance among higher education institution through empirical analysis with a sample of 304 students studying at five different universities in Somalia. Although model such as TAM and conducting research about ICTs readiness have reached relative maturity, this paper argues the need to assess the present state of ICTs readiness and acceptance in low income countries such as Somalia, and also its opportunities and challenges offered by ICTs applications within the higher education area should be well understood.

The results of this study lead to different crucial findings. Based on the research hypotheses, we found that with regards to ICT readiness among university students, the two determinant constructs of usefulness and ease of use had a significant and positive influence on ICT readiness acceptance. This means that in the Somali higher education context, the people who are used to ICT for teaching and learning activities may develop habitual use of technologies to contribute to quality of systems, because the users have believed that the ICT applications are easy to use and important for managing learning and teaching processes. These results are in agreement with previous researches on ICT acceptance such as the study on user's intention to adopt mobile learning readiness in higher education (Cheon et al., 2012; Mohammadi, 2015; Park et al., 2012; Yeap, Ramayah, & Soto-Acosta, 2016). Similarly, these results also supported other period studies concerning diversity settings, including m-commerce adoption (Teo, Cheah, Leong, Hew, & Shum, 2012), online banking (Tan, Chong, Ooi, & Chong, 2010), internet services adoption (Hsu & Yeh, 2017). This study also reported that perceived usefulness and an easy-to-use new technological interface could increase a user's collaboration, which is most helpful in their learning. Thus, students and learners of Somali higher education institutions would intend to use ICT tools to interact and communicate with their peers, and more if it has proven to be easy.

On the organizational level, the results of this study indicate that availability of infrastructure (H3) has positively influenced usefulness towards ICT readiness acceptance. The study shows that university students feel that support services (e.g. technologies such hardware, software, and networking) facilitate their academic life experiences. This finding is consistent with other studies (Aboelmaged, 2014), because availability of ICT infrastructure in a university signals technology readiness efforts and the meaningful use of ICT applications for their learning issues, which lead to improvements among university students to utilize new technologies (Aboelmaged, 2014; Birba & Diagne, 2012). Apart from the existence of some ICT infrastructure in universes, Table 4 shows results that indicate that the internet affordability (H4) and network speed and quality (H5) dimensions of organizational readiness have been negatively linked to perceived usefulness towards ICT readiness acceptance. Period research conducted by Penard et al. (2015) highlighted that internet network speed and affordability are two important factors which may enhance accessibility and use of internet in higher education. Thus, this study suggested that Somali universities should improve their internet access conditions which include infrastructure, high-speed internet network services, etc.

Concerning personal ICT readiness, four of research hypotheses was confirmed during analysis process, the Optimism (H6a &b), Innovativeness (H7a&b), Discomfort (H8a &b) and Insecurity (H9a &b). Our findings indicate that Optimism has a stronger positive relationship on perceived usefulness and perceived ease of use towards ICT readiness acceptance than others. This confirms that the results found by de Melo Pereira, Ramos, Gouvêa, and da Costa (2015) in their studies indicated that the construct of optimism in the model TRI has positively influenced the determinant constructs (usefulness and Ease of Use), coherently with other studies (Erdoğmuş & Esen, 2011). On the contrary, in the present study, the innovativeness (H7b), discomfort (H8b) and insecurity (H9b) showed no significant effects. Whereas H7a, H8a and H9a were supported accordingly. This finding is inconsistent with previous results by Kuo et al. (2013), who in their studies on technology readiness acceptance of mobile electronics, concluded that the innovativeness, discomfort and insecurity have a significant effect on Ease of Use but not on usefulness. This result is consistent with other findings (Erdoğmuş & Esen, 2011; Walczuch et al., 2007). It was also found that, when users believe that they can maintain their optimism and innovativeness, they feel more ready to use new technology. On the other hand, users feeling discomfort and insecurity indicate their inability to control the technology and worry about the negative consequences of the system or their lack of trust in the system (Aboelmaged, 2014).

Regarding bonds with individual control beliefs, this research reveals important results. ICT Self-efficacy (H10) and Teaching and Learning Autonomy (H11) have a significant and positive path to the perceived ease of use towards ICT readiness acceptance, which implies that individual learners' beliefs about capabilities and skills allow them to have sufficient control of new technologies for learning willingness. Results from Yeap et al. (2016) indicated that two constructs, CT Self-efficacy and learning autonomy, have a major influence on system acceptance. The current findings are also similar with what Cheon et al. (2012) found, which was that learning autonomy and self-efficacy have a significant relationship to ease of use. In other words, these constructs are compatible with perceived controls behavior, since the two constructs describe the facilitating conditions; whereas individuals' beliefs about using new ICT tools to conduct an activity, tend to have higher motivation to perform and control the process of learning.

**Conclusion, Limitations and Future Research**

To sum up, the purpose of this study was to investigate the factors that influence users' ICT readiness acceptance in Somalian higher education institutions by proposing a model based on the Technology Acceptance Model, with constructs driven from Technology readiness index (TRI) and existing literature. The model was validated by using the smart PLS-SEM approach. The results of this study showed the importance of Availability of ICT infrastructure, ICT Self-efficacy and Teaching and Learning Autonomy, along with TAM dimensions (Ease of use and Usefulness) in academics, when it comes to ICT readiness acceptance. The study was found that the Perceived Ease of Use is the predictor of ICT readiness acceptance ($\beta$=0.404) and explained 27% of the variance ($R^2$=0.273) in ICT readiness acceptance. Based on the results, the second strongest predictor in this study was Perceived Usefulness ($\beta$=0.314) with a 30% of the variance ($R^2$=0.306). A combination of factors, including Optimism, Innovativeness, Discomfort and Insecurity, showed mixed results. The present study contributed to the body of knowledge by providing valuable information and awareness to the higher education institutions on improving academic achievements.

As with all studies, this study has certain limitations which must be addressed in future research. Firstly, the demographic profiles under investigation are only senior students in higher educational institutions; university lecturers, staff and contributions from other parties of interest were not taken into consideration. Therefore, we recommend further research to examine perspectives of not only university students but also lecturers/instructors and staff, to check if their responses may have different results than the ones who have participated in this research. Secondly, since our study only focused on the education sector, the results of it cannot be regarded as a representation of other sector's ICT readiness acceptance in Somalia. Thus, to generalize the findings to other sectors such business, health or others, we suggest future research in different sectors. Thirdly, the survey of this study was restricted only to the unique environment of the Somali context. Further research should examine other developing countries as well, to compare if any discrepancies exist in the ICT readiness acceptance in higher educations.

## Acknowledgment

The authors thank the Center of Research and Development of SIMAD University, for funding this study under Research Grant Scheme (The use of ICT to support teaching and learning in higher education), with Grant No: SU-DA-RG-2018-005. The authors also thank all the respondents of this research.